\documentclass[12pt]{article}

\usepackage{geometry}
\usepackage[textwidth=8em,textsize=small]{todonotes}

\usepackage{graphicx}
\usepackage{natbib}
\usepackage{amsmath,amssymb,bm}
\usepackage{float}
\usepackage{cases}
\usepackage{pdflscape}
\usepackage{amsmath}
\usepackage{algorithm}
\usepackage{natbib}
\usepackage{url}
\usepackage{algpseudocode}
\usepackage{amssymb}
\usepackage{color}
\usepackage[colorlinks, linkcolor=blue, citecolor=blue]{hyperref}
\usepackage{etoolbox}

\newcommand{\abs}[1]{\left\vert#1\right\vert}
\newcommand{\dd}{\mathrm{d}}
\def\m{\mathcal}
\renewcommand{\hat}{\widehat}
\renewcommand{\bar}{\overline}

\title{Bayesian Nonparametric Nonhomogeneous Poisson Process with Applications to USGS Earthquake Data}

\author{Junxian Geng~~Wei Shi~~Guanyu Hu}

\begin{document}
\maketitle
\begin{abstract}
Intensity estimation is a common problem in statistical analysis of spatial point pattern data. This paper proposes a nonparametric Bayesian method for estimating the spatial point process intensity based on mixture of finite mixture (MFM) model. MFM approach leads to a consistent estimate of the intensity of spatial point patterns in different areas while considering heterogeneity. An efficient Markov chain Monte Carlo (MCMC) algorithm is proposed for our method. Extensive simulation studies are carried out to examine empirical performance of the proposed method. The usage of our proposed method is further illustrated with the analysis of the Earthquake Hazards Program of United States Geological Survey (USGS) earthquake data.\bigskip

\end{abstract}

\section{Introduction}\label{sec:intro}
Earthquake analysis is a widely discussed topic in the field of seismology and dates back as early as 1894 \citep{omori1894after}. Existing literatures \citep{schoenberg2003multidimensional,charpentier2015modeling,hu2018bayesian,Nas2019,yang2019bayesian} discussed spatial patterns and important covariates of the occurrence and the magnitude of earthquakes. Most earthquakes occur in seismic belt which is the narrow geographic zone on the Earth's surface. This spatial feature indicates the potential heterogeneity of the earthquake activities over the space. \cite{dasgupta1998detecting} considered the problem of detecting features, such as minefields or seismic faults, to show the heterogeneity of the earthquakes between sub-areas. Not only the earthquake activities but also locations of the tree species will show the heterogeneity pattern over space. For the data with heterogeneity features among sub-areas, we naturally consider it as a clustering problem in statistical analysis. 

Spatial point process model assumes that the randomness is associated with the locations of the points, which is a natural model for the earthquake data. It has been developed for analyzing spatial point pattern data \citep{moller2003statistical,diggle2013statistical}. A common problem in statistical analysis of spatial point patterns is to investigate the intensity of spatial point patterns. Traditional parametric estimation approaches are discussed in \cite{diggle2013statistical} and \cite{moller2003statistical}. \cite{baddeley2012nonparametric} described nonparametric (kernel and local likelihood) methods for estimating the effect of spatial covariates on the point process intensity. They assumed that the point process intensity is a function of the covariates, and studied nonparametric estimator of this function. In addition to the frequentist approaches, existing literatures also discussed the Bayesian approaches for spatial point process. \cite{leininger2017bayesian} proposed a full Bayesian model for estimating the intensity of spatial point process and considering model criticism and model selection both in-sample and out-of-sample. \cite{shirota2017approximate} proposed an approximate Bayesian computation (ABC) for the determinantal point process. While most existing literatures focus on intensity estimation of spatial point pattern, people pay less attention to clustering structure detection along with the intensity estimation.
 
Motivated by the features of the earthquake data and the limitations of the existing methods previously discussed, this paper introduces a Bayesian nonparametric estimation of nonhomogenous Poisson process to capture the heterogeneity pattern in the data. 

Bayesian inference provides a probabilistic framework for simultaneous inference of the number of clusters and the clustering configurations, although the case of unknown number of clusters poses computational burdens. In a fully Bayesian framework, complicated searching algorithms in variable dimensional parameter space such as the reversible jump MCMC algorithm \citep{green1995reversible}, assign a prior on the number of clusters which is required to be updated at each iteration of an MCMC algorithm.  Those algorithms are difficult to implement and automate, and are known to suffer from lack of scalability and mixing issues. 

Bayesian nonparametric approaches such as the Chinese restaurant process \citep[CRP;][]{pitman1995exchangeable} offer choices to allow uncertainty in the number of clusters. It has been empirically and theoretically observed that CRPs often have the tendency to create tiny extraneous clusters \citep{miller2018mixture}. We instead use the mixture of finite mixture (MFM) approach of \cite{miller2018mixture} which prunes the tiny extraneous clusters and, consequently, estimates the number of clusters consistently. The consistency on the number of clusters from MFM has been shown in \cite{miller2018mixture,rousseau2011asymptotic,geng2018probabilistic}. Moreover, the MFM model has a P\'{o}lya urn scheme similar to the CRP which is exploited to develop an efficient MCMC algorithm. In particular, we obtain an efficient Gibbs sampler by analytically marginalizing over the number of clusters and thus avoid complicated reversible jump MCMC algorithms or allocation samplers. 

The main contribution of this work is that we introduce a nonparametric Bayesian approach based on MFM for simultaneous inference of the intensity of spatial point pattern  and the clustering information (number of clusters and the clustering configurations) over the space. Furthermore, an efficient MCMC algorithm is proposed for our method based on Gibbs sampler. Our approach avoids sampling from complicated reversible jump MCMC algorithms or allocation samplers. In addtion, our proposed Bayesian approach reveals some interesting features of the earthquake data set.

The rest of the article is organized as follows. We start with a brief introduction of the data we use in Section \ref{usgs}. In addition, a review of the nonhomogeneous Poisson process (NHPP) model and MFM approach are discussed in Section \ref{sec:spp} and Section \ref{sec:MFM}, and then our Bayesian nonparametric intensity estimation model is proposed for nonhomogeneous Poisson process based on MFM in Section \ref{sec:hierachical_model}. Furthermore, we present priors and posteriors and develop a Markov chain Monte Carlo (MCMC) sampling algorithm in Section \ref{sec:bayes_comp}. Simulation studies and comparisons with existing methods are provided in Section \ref{sec:simu}. In Section \ref{sec:real_data}, the proposed method is employed to analyze the real data set of USGS. We conclude the article with some discussion in Section \ref{sec:discussion}. For ease of exposition, additional technical results are given in an appendix.

\section{USGS Earthquake Data}\label{usgs}
We consider the earthquake data from USGS, the Earthquake Hazards Program of United States Geological Survey (USGS), which can be accessed via \url{https://earthquake.usgs.gov/earthquakes/}, as our real data illustration. The dataset that we use for this analysis contains worldwide earthquakes which have magnitude over four from 10-01-2018 to 12-31-2018. This is mainly due to the belief that earthquakes with the magnitude above four will have some impacts on human daily life.   

\begin{figure}[tbp]
	\center
	\includegraphics[width= 2.5 in]{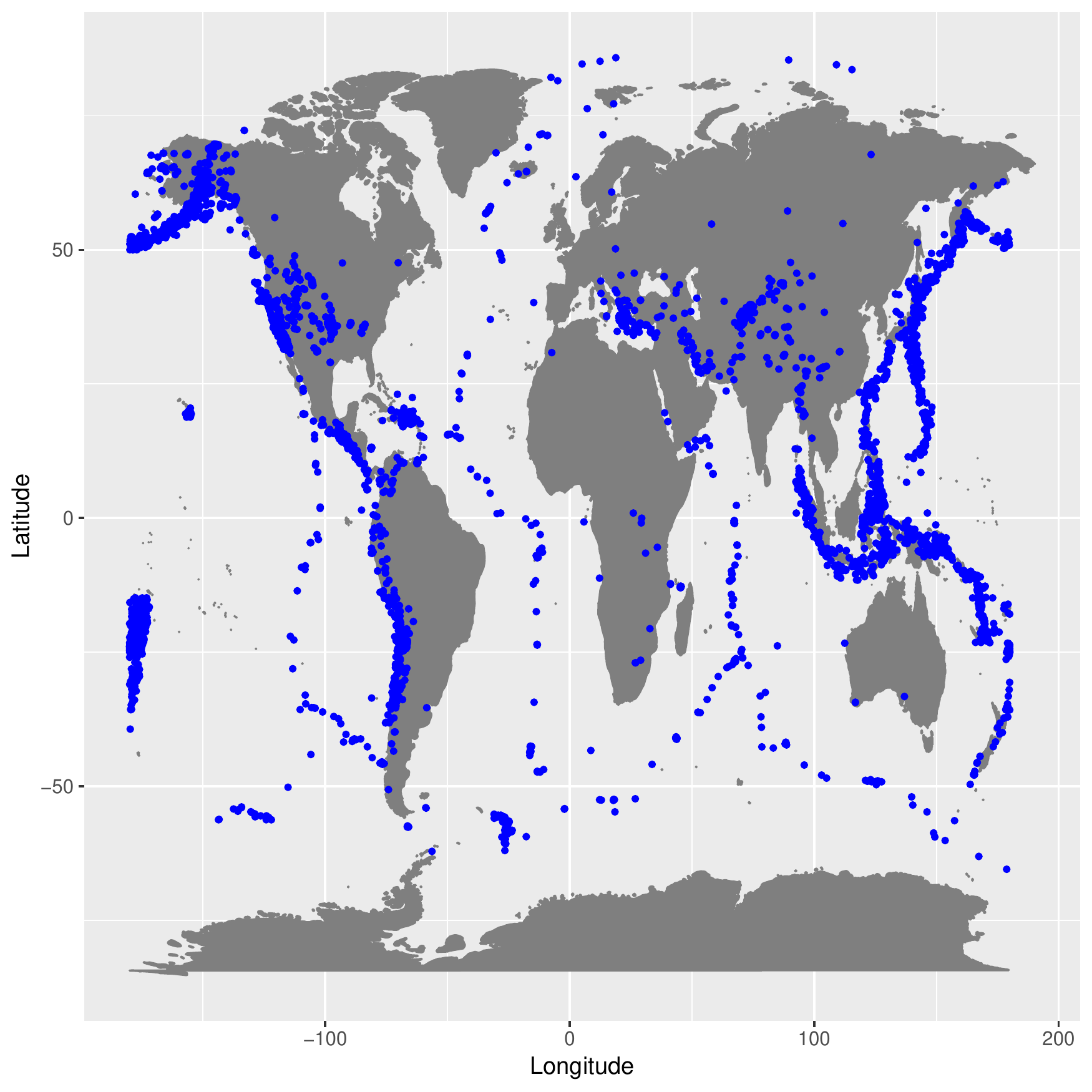}
	\includegraphics[width= 2.5 in]{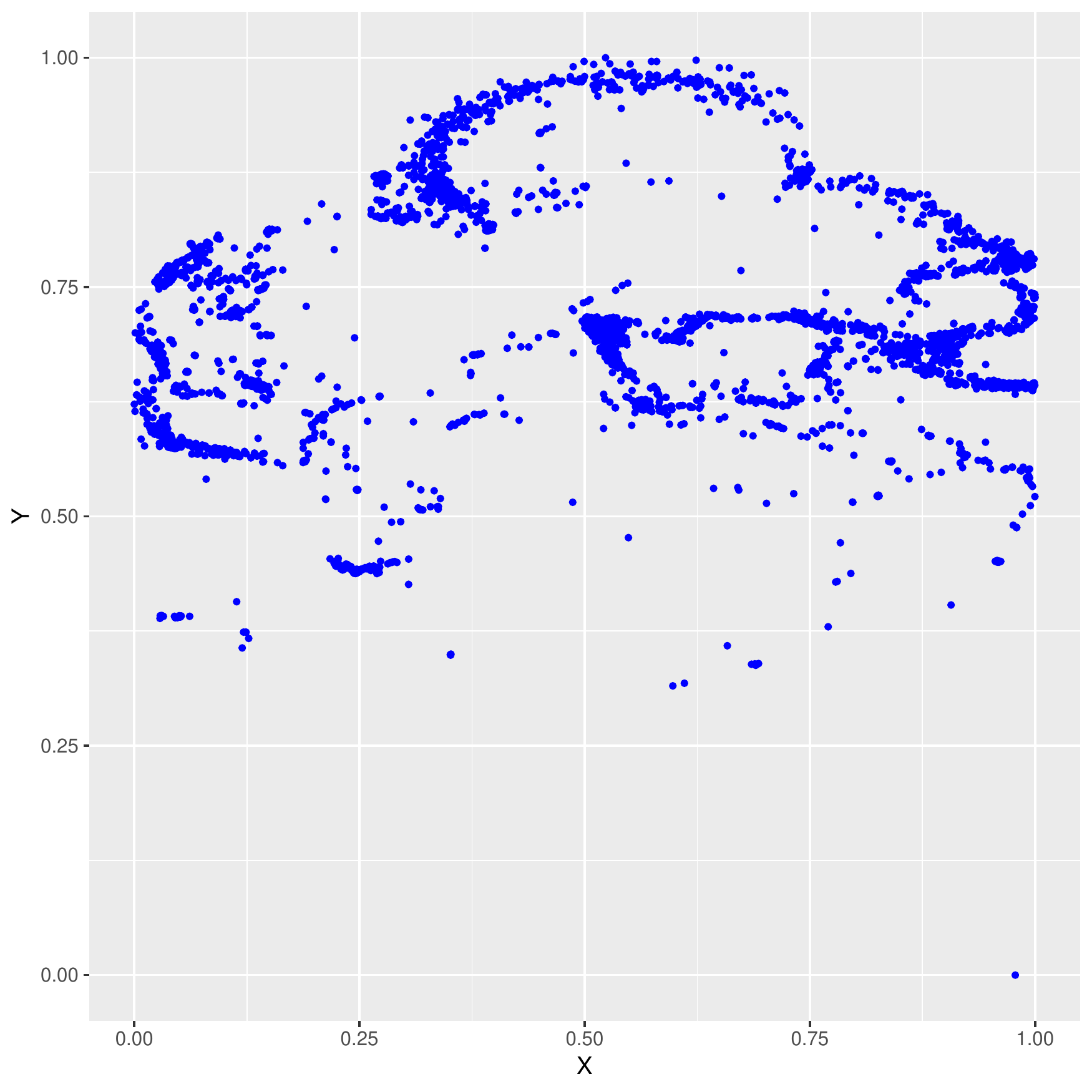}
		\caption{Map of Locations of the Earthquake (Left: Original Map; Right: Location in Unite Square)}
	\label{fig:earthquake_map}
\end{figure}

The total number of earthquakes in our dataset is 7701. The map of locations of the earthquakes we analyze is shown in the left panel of Figure \ref{fig:earthquake_map}. In order to analyze location relationship properly, we transform the latitude and longitude of the earthquakes to a $[0,1]\times [0,1]$ square. The locations of earthquake in a unit square are then shown in the right panel of Figure \ref{fig:earthquake_map}. From the left panel in Figure \ref{fig:earthquake_map}, we see that there are nearly 90\% of the world not having any occurrences of the earthquakes during the study period. For the North America region, most earthquakes occurred in Alaska's central coast, extending north to Anchorage and Fairbanks, and  the coast from British Columbia to the Baja California Peninsula, where the Pacific plate rubs against the North American plate. South America earthquakes stretch the length of the continent's Pacific border. For the Asia area, most earthquakes occurred in where the Australian plate wraps around the Indonesian archipelago and also in Japan. 

The observations above clearly indicate the heterogeneity of the earthquake activities over the space.
\section{Methodology}\label{sec:method}

\subsection{Spatial Poisson Process}\label{sec:spp}

A natural model for the earthquake data is a spatial point process model which assumes that the randomness is associated with the locations of the points. Let $\bm{y}=(s_1,s_2,...,s_\ell)$ be the set of locations for points that are observed in a bounded region $\mathcal{B}\subseteq \mathcal{R}^2$, which is a realization of spatial point process $\bm{Y}$. This is called a spatial point pattern.  The process $N_{\bm{Y}}(A)=\sum_{i=1}^\ell 1(s_i\in A)$ is a counting process associated with the spatial point process $\bm{Y}$,
 which counts the number of points of $\bm{Y}$ for area $A\subseteq\mathcal{B}$.
 For the process $\bm{Y}$, there are many parametric distributions for a finite set of count variables like Poisson processes, Gibbs processes, and Cox processes \citep{diggle2013statistical}. In this work, we focus on spatial Poisson processes.
For the Poisson process $\bm{Y}$ over $\mathcal{B}$ with intensity function $\lambda(\bm{s})$, $N_{\bm{Y}}(A)\sim \text{Poisson}(\lambda(A))$, where $\lambda(A)=\int_A\lambda(s) \dd s$. In addition, if two areas $A_1$ and $A_2$ are disjoint, then $N_{\bm{Y}}(A_1)$ and $N_{\bm{Y}}(A_2)$ are independent, where $A_1 \subseteq \mathcal{B}$ and $A_2 \subseteq \mathcal{B}$.
  Based on properties of the Poisson process, we obtain $E(N_{\bm{Y}}(A))=\text{Var}(N_{\bm{Y}}(A))=\lambda(A)$. When $\lambda(s)=\lambda$, we have constant intensity over the space $\mathcal{B}$, and in this special case, $\bm{Y}$ reduces to a homogeneous Poisson process (HPP). In more general cases, $\lambda(s)$ can be spatially varying, which leads to a nonhomogeneous Poisson process (NHPP). For the NHPP, the likelihood on $\mathcal{B}$ is given by
\begin{align}
	L=\frac{\prod_{i=1}^\ell \lambda(s_i)}{\exp(\int_{\mathcal{B}} \lambda(s) \dd s)},
	\label{pploglike}
\end{align}
where $\lambda(s_i)$ is the intensity function for location $s_i$.

\subsection{Nonparametric Bayesian methods in Spatial Poisson Process}\label{sec:MFM}
In order to simplify the problem induced by non-homogeneity on intensity values, a common approach provided by \cite{teng2017bayesian} is to divide the spatial area into $n$ disjoint sub-areas such that we can make the assumption that the intensity is constant over each sub-area. 

A commonly used approach to simplify the problem induced by non-homogeneous intensity is to partition a spatial area into $n$ disjoint sub-areas and assume constant intensity over each sub-srea \citep{teng2017bayesian}. The number $n$ is usually referred to as the pixel resolution or partition number over the space.t Let $A_1, A_2, \cdots, A_n$ be a partition of $\mathcal{B}$, i.e., they are disjoint
subsets such that $\bigcup_{i=1}^n A_i = \mathcal{B}$. For each region $A_i, i=1,\cdots,n$, we have constant intensity $\lambda_i$ over region $A_i$. The likelihood in \eqref{pploglike} is written as:
\begin{equation}
	\label{eq:papprox}
	L=\prod_{i=1}^n f_{\text{poisson}}(N_{\bm{Y}}(A_i)|\lambda_i),
\end{equation} 
where $f_{\text{poisson}}$ is the probability mass function of the Poisson distribution. In later sections, we use $N(A_i)$ to denote $N_{\bm{Y}}(A_i)$.

A latent clustering structure provides the ability to accommodate the heterogeneity on intensity values for each sub-area $\lambda_i$. Let $\m Z_{n, k} = \big\{(z_1, \ldots, z_n) : z_i \in \{1, \ldots, k\}, 1 \le i \le n \big\}$ denote all possible clusterings of $n$ sub-areas into $k$ clusters, where $z_i \in \{1, \ldots, k\}$ denotes the cluster assignment of the $i$th sub-area. There are two problems to solve: the number of clusters and the cluster assignment for each sub-area.

Under the frequentist framework, a two-stage procedure can be implemented where 
we first estimate the number of clusters and then estimate the cluster assignments based on the cluster number. Such two stage procedures may ignore uncertainty of the estimation of the number of clusters in the first stage, and are prone to increased erroneous cluster assignments
in the second stage.

In contrast, Bayesian models offer a natural solution to simultaneously estimate the number of clusters and cluster assignments. The Chinese restaurant process \citep[CRP;][]{pitman1995exchangeable, neal2000markov} offers choices to allow for uncertainty in the number of clusters by assigning a prior distribution on $(z_1, z_2, \ldots, z_n)$. In the CRP, $z_i, i=2, \ldots, n$ are defined through the following conditional distribution  \citep[i.e., a P\'{o}lya urn scheme,][]{blackwell1973ferguson} 
\begin{eqnarray}\label{eq:crp}
P(z_{i} \mid z_{1}, \ldots, z_{i-1})  \propto   
\begin{cases}
\abs{c}  , \quad  \text{at an existing cluster labeled}\, c\\
\alpha,  \quad \quad \quad \, \text{at a new cluster}.  
\end{cases}
\end{eqnarray}
Here $\abs{c}$ refers to the size of cluster labeled $c$, and $\alpha$ is the concentration parameter of the underlying Dirichlet process. 
At time $n = 1$, the trivial partition $\{ \{1\} \}$ is obtained with probability $1$. At time $n + 1$, the $n + 1$th element is either  i) added to one of the blocks of the partition $\mathcal{C}_n$, where each block is chosen with probability  $\abs{c}/(n + \alpha)$, or ii) added to the partition $\mathcal{C}_n$ as a new singleton block, with probability $\alpha/(n + \alpha)$. Here~$\mathcal{C}_n$ denotes a partition of the set $\{1, 2, 3, \ldots,n\}$.  
Let $t = \abs{\mathcal{C}_n}$ denote the number of blocks in the partition $\m C_n$. Under \eqref{eq:crp}, one can obtain the probability of block-sizes $\bm{b}= (b_1, b_2, \ldots, b_t)$ of a partition $\mathcal{C}_n$ as
\begin{eqnarray} \label{eq:sizeprobcrp}
p_{\mathrm{DP}}(\bm{b}) \propto  \prod_{j=1}^t  b_j^{-1}.  
\end{eqnarray}
It is clear from \eqref{eq:sizeprobcrp} that the CRP assigns large probabilities to clusters with relatively smaller sizes, which results in producing extraneous clusters in the posterior leading to inconsistent estimation on the {\em number of clusters} even when the sample size goes to infinity. \cite{miller2018mixture} proposed a modification to the CRP, which is called a mixture of finite mixtures (MFM) model, to circumvent this issue:   
\begin{eqnarray}\label{eq:MFM}
\begin{split}
k & \sim p(\cdot), \\
(\pi_1, \ldots, \pi_k) \mid k &\sim \mbox{Dir}(\gamma, \ldots, \gamma), \\
 z_i \mid k, \pi & \sim \sum_{h=1}^k  \pi_h \delta_h, \quad  i=1, \ldots, n, 
\end{split}
\end{eqnarray}
 where $p(\cdot)$ is a proper probability mass function on $\{1, 2, \ldots\}$, and $\delta_h$ is a point-mass at $h$.  
\cite{miller2018mixture} showed that the joint distribution of $(z_1, \ldots, z_n)$ under \eqref{eq:MFM} admit a P\'{o}lya urn scheme akin to the CRP:
\begin{eqnarray}\label{eq:mcrp}
P(z_{i}\mid z_{1}, \ldots, z_{i-1})  \propto   
\begin{cases}
\abs{c} + \gamma  , \quad  \text{at an existing cluster labeled}\, c\\
\frac{V_n(t+1)}{V_n(t)} \gamma,  \quad \quad \quad \, \text{at a new cluster}, 
\end{cases}
\end{eqnarray}
where $V_n(t)$ is a coefficient of partition distribution that need to be precomputed, 
\begin{align*} 
\begin{split}
V_n(t) &= \sum_{k=1}^{+\infty}\dfrac{k_{(t)}}{(\gamma k)^{(n)}} p(k)
\end{split}					
\end{align*} 
where $k_{(t)}=k(k-1)...(k-t+1)$, and $(\gamma k)^{(n)} = {\gamma k}(\gamma k+1)...(\gamma k+n-1)$. (By convention, $x^{(0)} = 1$ and $x_{(0)}=1$).
Compared to the CRP, the introduction of new clusters is slowed down by a factor $V_n(\abs{\mathcal{C}_{n-1}}  +1)/ V_n(\abs{\mathcal{C}_{n-1}})$, thereby pruning the tiny extraneous clusters.

An alternative way to understand the natural pruning of extraneous clusters is through the probability distribution induced on the block-sizes $\bm{b} = (b_1, b_2, \ldots, b_t)$ of a partition $\mathcal{C}_n$  with $t= |\m C_n|$ under MFM.  In contrast to \eqref{eq:sizeprobcrp},  the probability of the cluster sizes $(b_1, \ldots, b_t)$ under the MFM is 
\begin{eqnarray} \label{eq:sizeprob}
p_{\mathrm{MFM}}(\bm{b}) \propto  \prod_{j=1}^t  b_j^{\gamma-1}.  
\end{eqnarray} 
From \eqref{eq:sizeprobcrp} and \eqref{eq:sizeprob}, it is easy to see that MFM assigns comparatively smaller probabilities to highly imbalanced cluster sizes.  The parameter~$\gamma$ controls the relative size of the clusters;  small $\gamma$ favors lower entropy~$\pi$'s, while large $\gamma$ favors higher entropy $\pi$'s.     

\subsection{MFM for Spatial Poisson Process}\label{sec:hierachical_model}
Adapting the MFM to the NHPP setting, the proposed model and prior can be expressed hierarchically as: 
\begin{align}\label{eq:MFMNHPP}
\begin{split}
& k \sim p(\cdot), \text{where $p(\cdot)$ is a p.m.f on } \{1,2, \ldots\}\\
& \lambda_r \stackrel{\text{ind}} \sim \mbox{Gamma}(a, b),  \quad r= 1, \ldots, k, \\
& P(z_i = j \mid \pi, k) = \pi_j, \quad j = 1, \ldots, k, \, i = 1, \ldots, n,  \\
& \pi \mid k \sim \mbox{Dirichlet}(\gamma, \ldots, \gamma),\\
& N(A_i) \mid z, \lambda, k \stackrel{\text{ind}} \sim \mbox{Poisson}(\lambda_{z_i}), \quad i = 1, \ldots, n, 
\end{split}
\end{align}
where $n$ is the number of areas in the sample space, $k$ is the number of clusters and $N(A_i)$ is the number of points in area $A_i$. A default choice of $p(\cdot)$ is a $\mbox{Poisson}(1)$ distribution truncated to be positive \citep{miller2018mixture}, which is assumed through the rest of the paper. We refer to the hierarchical model above as MFM-NHPP. 

\section{Computation}\label{sec:bayes_comp}
\subsection{The MCMC Sampling Schemes}\label{sec:mcmc}
Our goal is to sample from the posterior distribution of the unknown parameters $k$, $z = (z_1, \ldots, z_n) \in \{1, \ldots, k\}^n$ and $\lambda = (\lambda_1, \ldots, \lambda_n)$. The sampler is presented in Algorithm \ref{algorithm}, which efficiently cycles through the full conditional distributions of $z_i \mid z_{-i}$ for $i=1, 2, \ldots, n$ and $\lambda$, where~$z_{-i} = z \backslash \{z_i\}$. The details of the full conditional distributions are in Appendix \ref{FCD}. The main trick in the MFM approach \citep{miller2018mixture} for clustering is to analytically marginalize over the distribution of $k$ and exploit the P\'{o}lya urn scheme to develop an efficient Gibbs sampler. The marginalization over $k$ can avoid complicated reversible jump MCMC algorithms or even allocation samplers.

\begin{algorithm}[tbp]
\caption{Collapsed sampler for MFM-NHPP}
\label{algorithm}
\begin{algorithmic}[1]
\Procedure{c-MFM-NHPP} {}
\\ Initialize $z = (z_1, \ldots, z_n)$ and $\lambda = (\lambda_1, \ldots, \lambda_k)$.
\For{each iter $=1$ to $\mbox{M}$ }
\\ Update $\lambda = (\lambda_1, \ldots, \lambda_k)$ conditional on $z$ in a closed form as
\begin{align*} 
\begin{split}
\lambda_{r} \mid N,z & \sim \mbox{Gamma}(\bar{N}_{r}+a,n_{r}+b)
\end{split}			
\end{align*} 
Where $\bar{N}_{r}=\sum_{z_i=r} N(A_i)$, $n_{r} = \sum_{i=1}^{n}I(z_i=r),  r=1,\ldots,k$.  Here $k$ is the number of clusters formed by current $z$.
\\ Update $z = (z_1, \ldots, z_n)$ conditional on $\lambda = (\lambda_1, \ldots, \lambda_k)$, for each $i$ in $(1,...,n)$, we can get a closed form expression for $P(z_i = c \mid z_{-i}, N, \lambda)$:
\[\propto \left\{
\begin{array}{ll}
      [\abs{c} + \gamma] \mbox{ } \mbox{dPoisson}(N(A_i);\lambda_c) & \text{at an existing cluster c} \\
      \frac{V_n(\abs{\mathcal{C}_{-i}}  +1)}{V_n(\abs{\mathcal{C}_{-i}}} \gamma  m(N(A_i)) & \text{if c is a new cluster} \\
\end{array} 
\right. \]
where $\mathcal{C}_{-i}$ denotes the partition obtained by removing $z_i$ and 
\small
\begin{eqnarray*}
m(N(A_i)) =  \dfrac{b^a \Gamma(N(A_i)+a)}{\Gamma(a)(b+1)^{N(A_i)+a}N(A_i)!}.
\end{eqnarray*}
\normalsize
\EndFor
\EndProcedure
\end{algorithmic}
\end{algorithm}
 
\subsection{Inference of MCMC results}\label{sec:sum_mcmc}
The estimated parameters including cluster assignment $z$, intensities $
\lambda$ are determined for each replicate from the best post burn-in iteration selected using the Dahl's method \citep{Dahl:2006}.

\cite{Dahl:2006} proposed a least-squares model-based clustering for estimating the clustering of observations using draws from a posterior clustering distribution. In this method, we need to get the membership matrices for each iteration as $B^{(1)},...,B^{(M)}$, in which $M$ is the number of posterior samples obtained after burn-in iterations. Membership matrix $B$ is defined as:
\begin{align}
\begin{split}
B = (B(i,j))_{i,j\in \{1:n\}} = (z_i = z_j)_{n\times n},
\end{split}
\end{align}
where $B(i,j) \in \{0,1\}$ for all $i,j = 1,...,n$, and $B(i,j)=1$ means observations $i$ and $j$ are in the same cluster in a certain iteration. Then we calculate the least squares distance to Euclidean mean for each MCMC iteration and choose the the best of these iterations. The procedure can be described as below:
\begin{itemize}
\item
Calculate the Euclidean mean for all membership matrices $\bar{B} = \frac{1}{M} \sum_{m=1}^M B^{(m)}$.
\item
Find the iteration that has the least squares distance to $\bar{B}$ as:
\begin{align}
\begin{split}
C_{LS} = \text{argmin}_{m \in (1:M)} \sum_{i=1}^n \sum_{j=1}^n (B(i,j)^{(m)} - \bar{B}(i,j))^2
\end{split}
\end{align}
\end{itemize}
The least-squares clustering has the advantage that it uses information from all the clusterings via the pairwise probability matrix and is intuitively appealing because it selects the “average” clustering instead of forming a clustering via an external, ad hoc clustering algorithm.

\subsection{Model Assessment}\label{sec:model_comparision}

In this section, we discuss two model assessment criteria for spatial point process model. 
First, we introduce an in-sample model assessment criteria to assess the intensity fitness of point process model. Let $A_1, A_2, \cdots, A_n$ be a partition of $\mathcal{B}$, i.e., disjoint subsets such that $\bigcup_{i=1}^n A_i = \mathcal{B}$. The mean absolute error (MAE) is defined as:
\begin{equation}
	\text{MAE}=\frac{1}{n}\sum_{i=1}^n|\hat{\lambda}(A_i)-N(A_i)|,
\end{equation} 
where $\hat{\lambda}(A_i)$ is the estimated intensity of the region $A_i$ and $N(A_i)$ is the observed points of the region $A_i$. Under the model assessment framework, the model with smaller MAE value has better fitness.

The aim of the second criterion, logarithm of the Pseudo-marginal likelihood \cite[LPML;][]{gelfand1994bayesian}, is to evaluate the region resolution in our Bayesian nonparametric estimation. The LPML is defined as
\begin{align}
\label{eq:defLPML}
\text{LPML} = \sum_{i=1}^{n} \text{log}(\text{CPO}_i),
\end{align}
where $\text{CPO}_i$ is the conditional predictive ordinate (CPO) for the $i$-th subject. Based on the leave-one-out-cross-validation, the CPO estimates the probability of observing $y_i$ in the future
after having already observed $y_1,\cdots,y_{i-1},y_{i+1},\cdots,y_n$. The CPO for the $i$-th subject is defined as
\begin{align}
	\text{CPO}_i=f(y_i|\bm{y}_{-i}) \equiv \int f(y_i|\bm{\theta})\pi(\bm{\theta}|\bm{y}_{-i})d\bm{\theta},
	\label{cpo def}
\end{align}
where $\bm{y}_{-i}$ is shorthand for $\{y_1,\cdots,y_{i-1},y_{i+1}\cdots,y_n\}$,
\begin{align}
	\pi(\bm{\theta}|\bm{y}_{-i})=\frac{\prod_{j\neq i}f(y_j|\bm{\theta})\pi(\bm{\theta})}{c(\bm{y}_{-i})},
\end{align}
and $c(\bm{y}_{-i})$ is the normalizing constant.
The $\text{CPO}_i$ in \eqref{cpo def} can be expressed as
\begin{align}
	\text{CPO}_i=\frac{1}{\int\frac{1}{f(y_i|\bm{\theta})}\pi(\bm{\theta}|\bm{y}_{-i})d\bm{\theta}}.
	\label{CPO1}
\end{align}

Based on \citet{hu2019bayesianmodel}, a natural Monte Carlo estimate of the LPML
is given by
\begin{equation}
  \widehat{\text{LPML}}
= \sum_{j=1}^\ell\log \widetilde{\lambda}(s_j)-\int_{\mathcal{B}}\overline{\lambda}(u)\,du,
\label{eq:lpml_estimate}
\end{equation}
where
$\widetilde{\lambda}(s_j)=(\frac{1}{B}\sum_{b=1}^B\lambda(s_j|\bm{\theta_b})^{-1})^{-1}$,
$\overline{\lambda}(u)=\frac{1}{B}\sum_{b=1}^B
\lambda(u|\bm{\theta_b}) $, and $\{\theta_1,\theta_2,\cdots,\theta_B\}$ is a posterior sample.
In real data analysis, we do not know the true resolution of spatial domain. Based on the LPML in \eqref{eq:lpml_estimate}, we can evaluate the performance of different resolutions. The model with larger LPML value is favored.

\section{Simulation}\label{sec:simu}
\subsection{Simulation Setup}
We use simulation studies to illustrate the performance of proposed MFM-NHPP approach from multiple perspectives. The data generation process is described below, and will be followed for the rest of the section. \smallskip \\
\underline{{\bf Step 1:}} Fix the number of areas $n$ \& the true number of clusters $K$. \\
\underline{{\bf Step 2:}} Generate the true clustering configuration $z_0 = (z_{01}, \ldots, z_{0n})$ with $z_{0i} \in \{1, \ldots, K\}$. \\
\underline{{\bf Step 3:}} Construct the $\sqrt{n}\times\sqrt{n}$ intensity matrix $Q$; each term in the matrix has an intensity value from $\lambda = (\lambda_1, \ldots, \lambda_K)$. The intensity values $\lambda$ will vary in different scenarios.\\
\underline{{\bf Step 4:}}  Generate the number of points in each area $N(A_i) \sim \mbox{Poisson}(Q_{z_{0i}})$ independently for $1 \leq i \leq n$.

\begin{figure}[ht]
	\center
	\includegraphics[width= 2.2 in, height = 2 in]{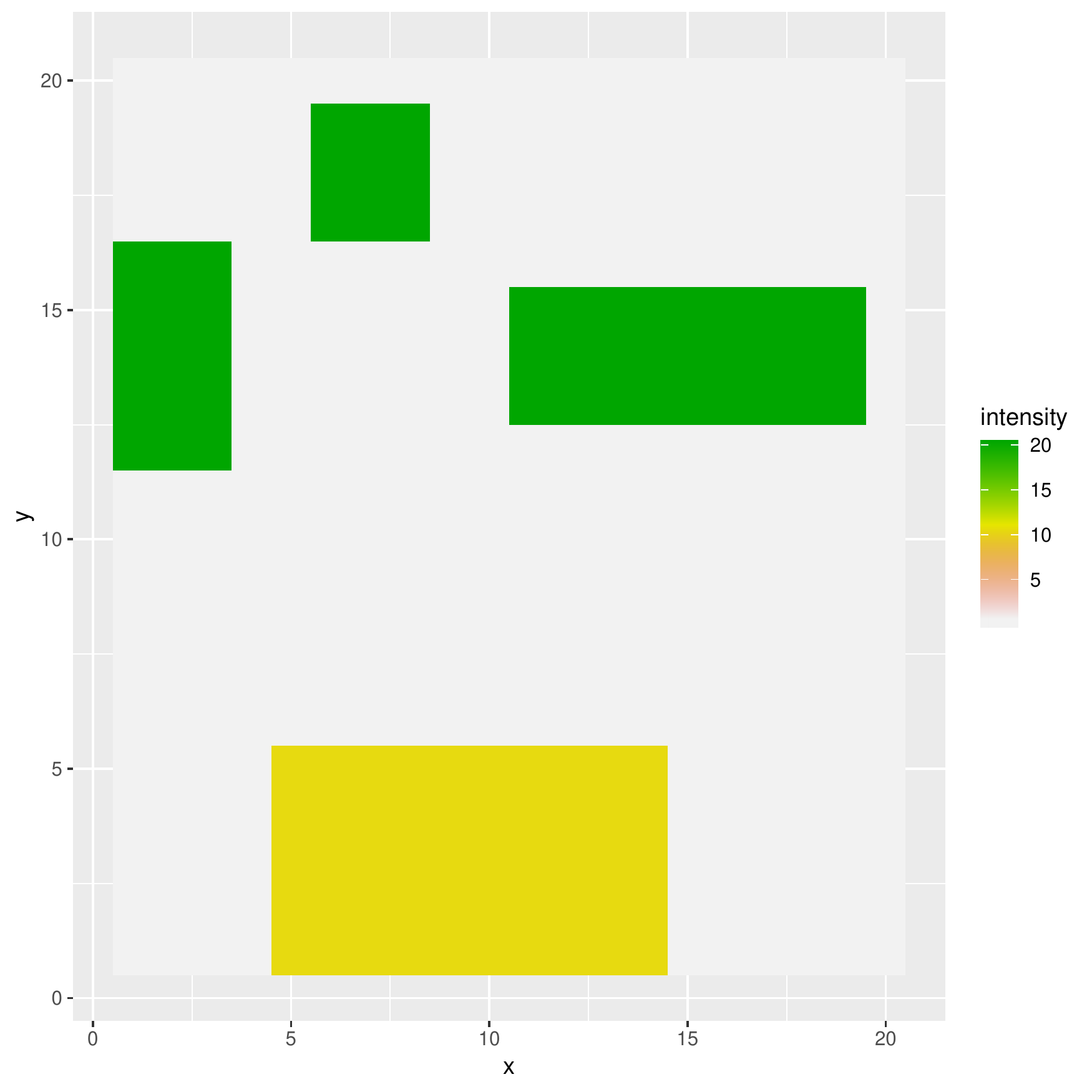}
	\includegraphics[width= 2 in, height = 2 in]{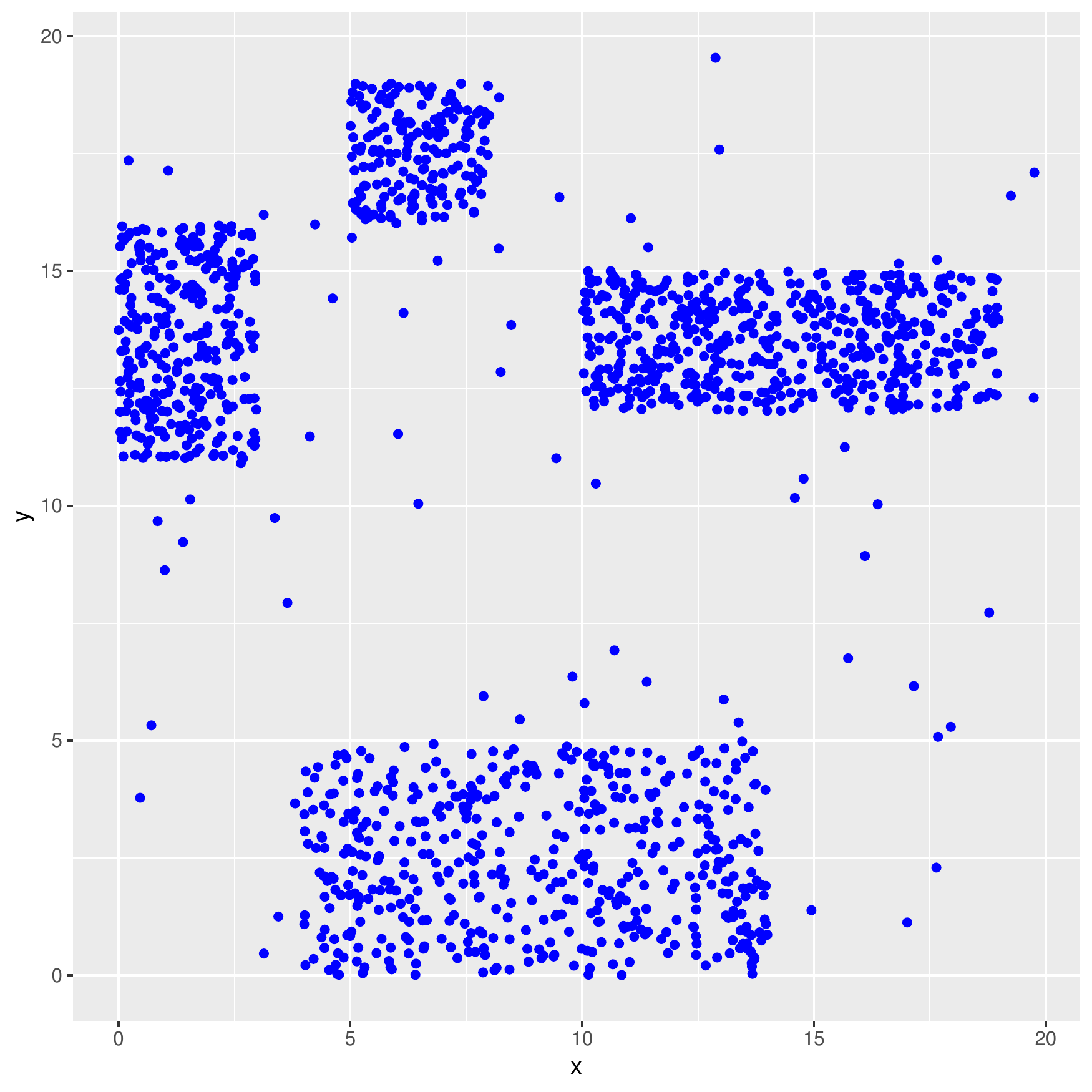}
	\caption{Scenario 1 intensity image (left) and simulated data (right)}\label{Simu1_data}
\end{figure}

In the simulation study, two different scenarios are considered. In the first scenario, we choose three different clusters with intensities $(0.2,10,20)$. The intensity image and one simulated data are shown in Figure~\ref{Simu1_data}.

\begin{figure}[ht]
	\center
	\includegraphics[width=2.2 in, height = 2 in]{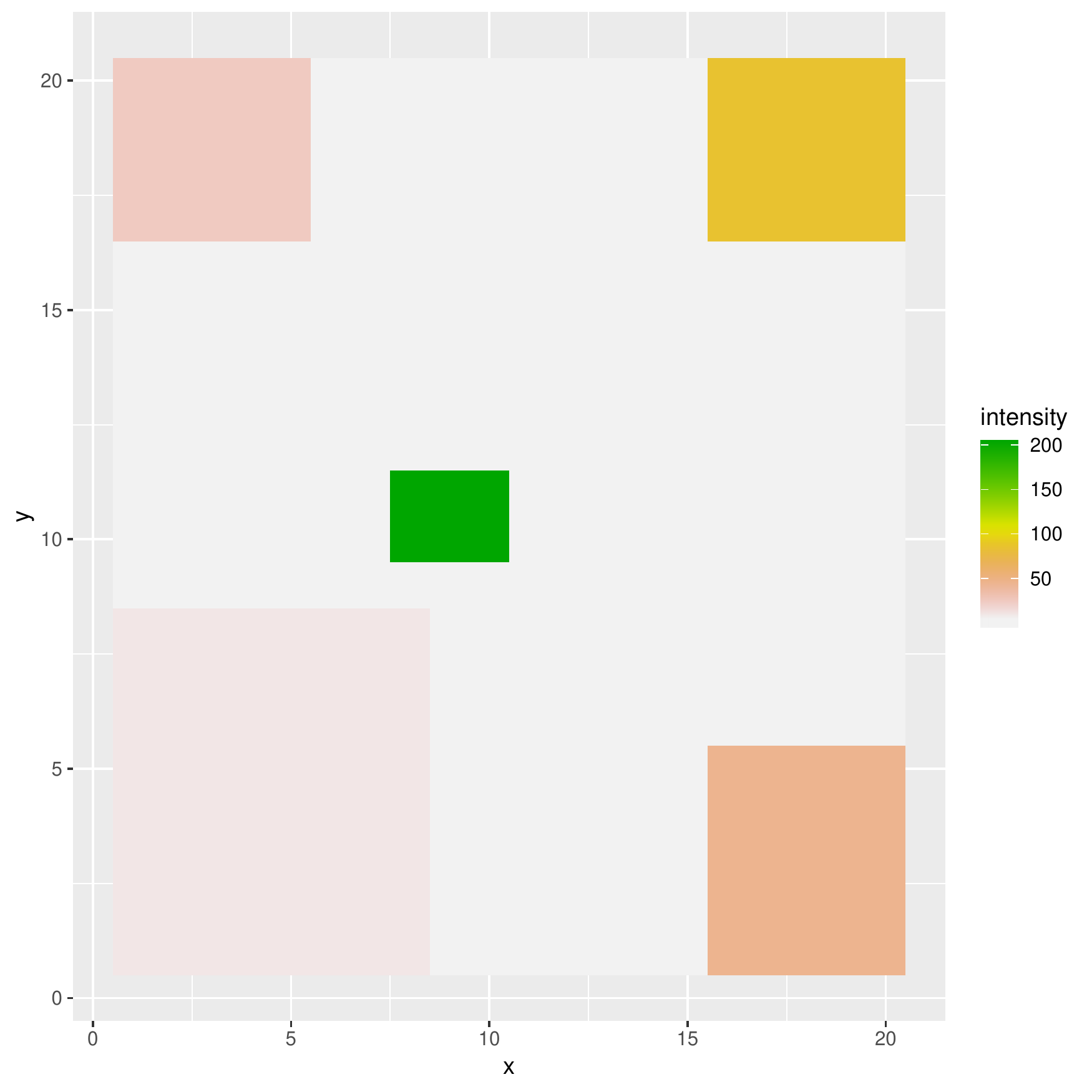}
	\includegraphics[width=2 in, height = 2 in]{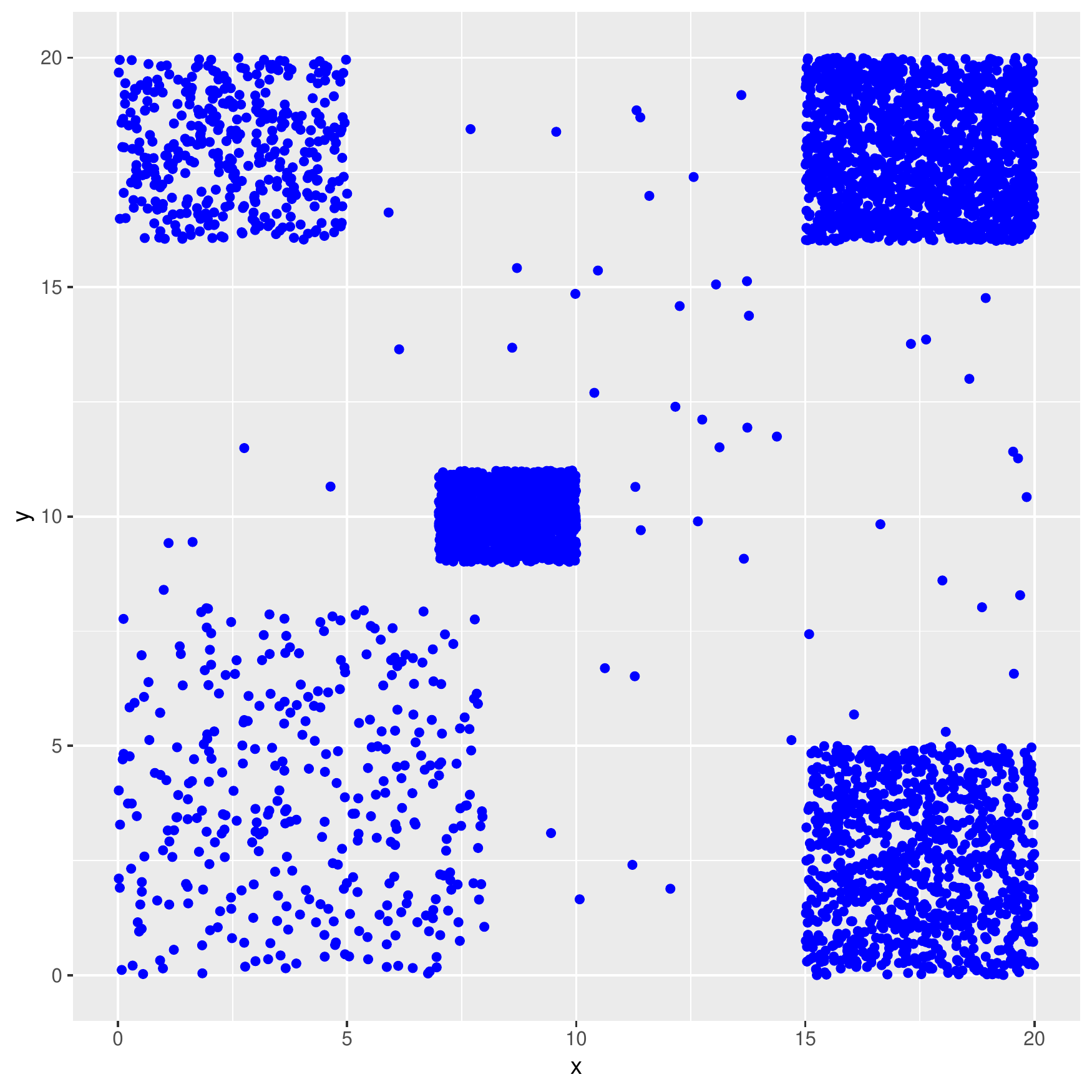}
	\caption{Scenario 2 intensity image (left) and simulated data (right)}
	\label{fig:scenario2}	
\end{figure}

In the second scenario, we choose six different clusters with intensities $(0.2,5,20,40,80,200)$. The intensity image and one simulated data are shown in Figure \ref{fig:scenario2}.

The first measure we are interested from the posterior is the estimation of $k$. The number of clusters $k$ is marginalized out in our collapsed Gibbs sampler, hence we do not directly obtain samples from the posterior distribution of $k$. However, $k$ can still be estimated based on the posterior distribution of $|z|$, the number of unique values (occupied components) in $(z_1, \ldots, z_n)$. We obtain $M$ posterior samples and obtain posterior summary measures based on samples post burn-in. Inference on the number of clusters and clustering configurations is obtained employing the modal clustering method of \cite{Dahl:2006}.

The second measure used in our performance evaluation is the Rand index \citep{rand1971objective}, which can be used to measure the accuracy of clustering. The Rand index $\mathrm{RI}$ is defined as 
\begin{eqnarray*}
 \mathrm{RI} = \frac{a+b}{a+b+c+d} = \frac{a+b}{{n \choose 2 }}, 
\end{eqnarray*}
where $\mathcal{C}_1 = \{X_1, \ldots, X_r\}$ and $\mathcal{C}_2 = \{Y_1, \ldots, Y_s\}$ are two partitions of $\{1, 2, \ldots, n\}$, and $a, b, c$ and $d$ respectively denote the number of pairs of elements of $\{1, 2, \ldots, n\}$ that are (a) in a same set in $\m C_1$ and a same set in $\m C_2$, (b) in different sets in $\m C_1$ and different sets in $\m C_2$, (c) in a same set in $\m C_1$ but in different sets in $\m C_2$, and (d) in different sets in $\m C_1$ and a same set in $\m C_2$. RI ranges from 0 to 1 with a higher value indicating a better agreement between the two partitions. In particular, $\mathrm{RI} = 1$ indicates $\m C_1$ and $\m C_2$ are identical (modulo labeling of the nodes). 

Another group of measures from the posterior are the estimation of intensity values for each cluster $\lambda = (\lambda_1, \ldots, \lambda_k)$. We report two types of estimations, one is based on the posterior sample in the iteration chosen by the modal clustering method of \cite{Dahl:2006}, and the other one is based on posterior mean post burn-in.

Without loss of generality, in all the simulation examples considered below, we employ Algorithm \ref{algorithm} with $n = 400~(20\times20)$, $\gamma =1$ and $a=b=1$ to fit the MFM-NHPP model; and we refer to this as the MFM-NHPP algorithm.  A truncated Poisson prior with mean $1$ is assumed on~$k$. The initial number of clusters is set to $5$, and we randomly allocate the cluster configurations in all the examples. The initial values for $\lambda = (\lambda_1, \ldots, \lambda_k)$ are from the prior distribution. We experiment with various other choices and do not find any evidence of sensitivity to the initialization.

\subsection{Convergence Diagnostics}\label{sec:conv}
\begin{figure}[tbp]
\centering
	\includegraphics[width=2 in]{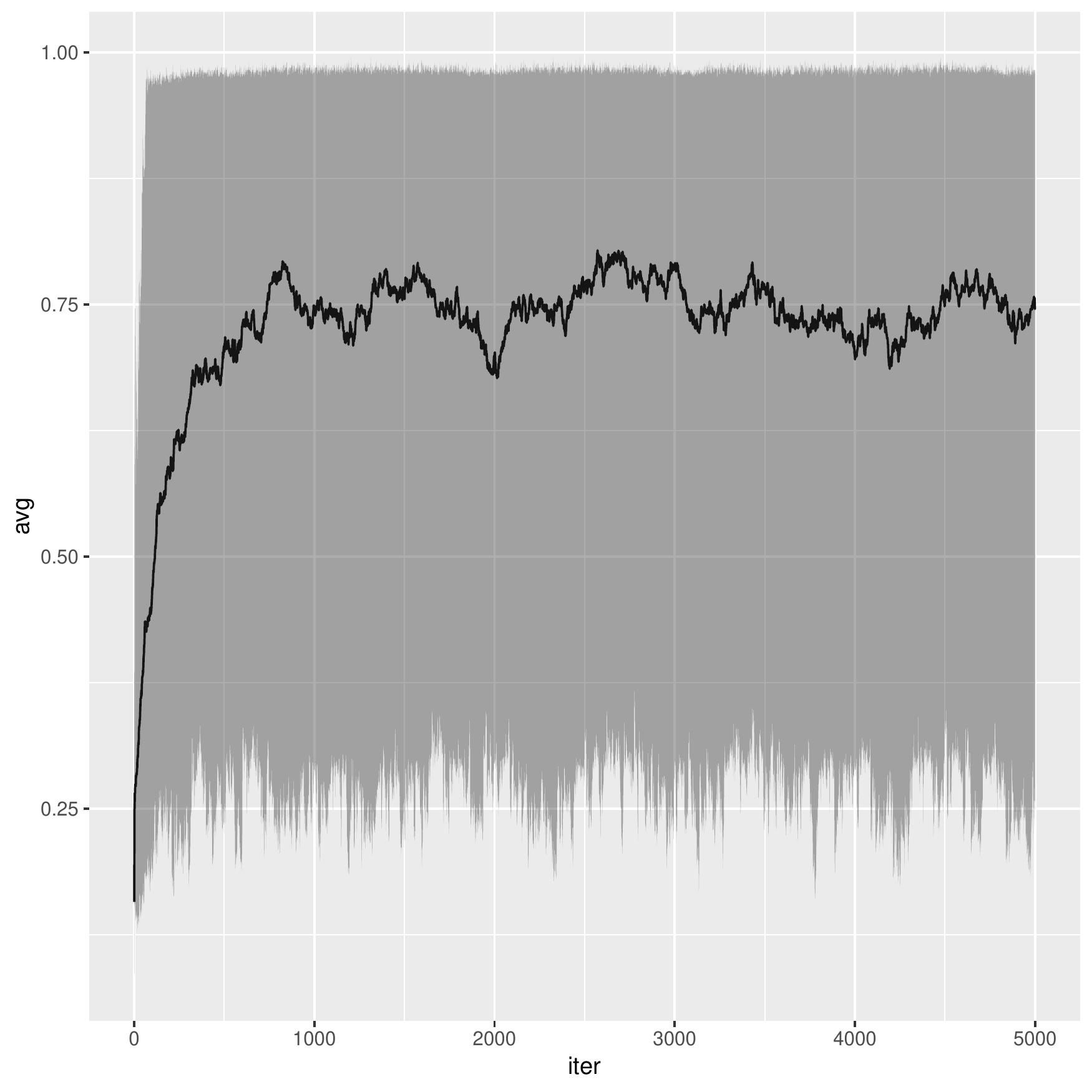}
	\includegraphics[width=2 in]{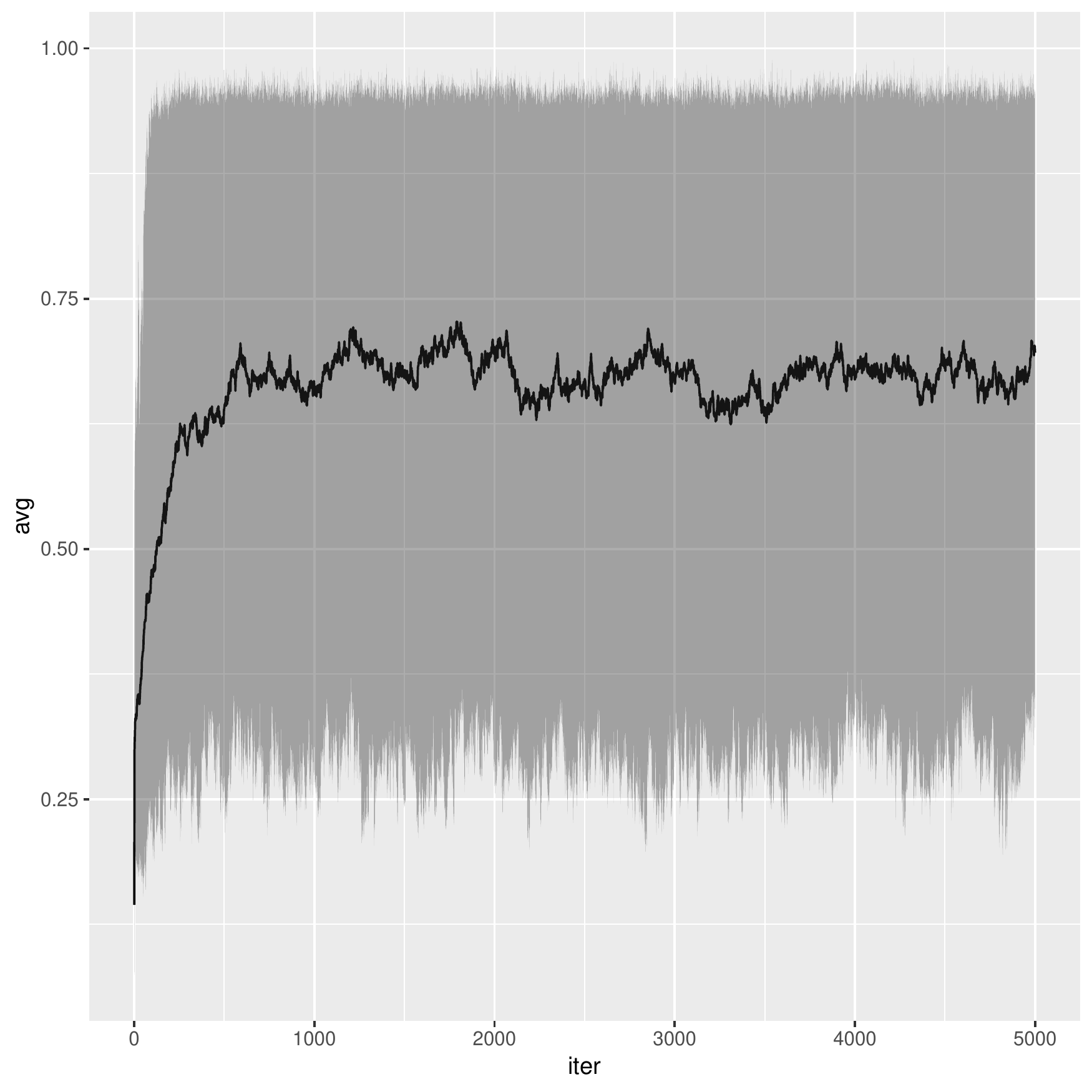}
	\caption{Average Rand Index for scenario 1 (left) and scenario 2 (right)}
	\label{fig:randindx}
\end{figure}

We present average value of $\mathrm{RI}(z, z_0)$ for the first 5000 MCMC iterations from 100 randomly chosen starting configurations for the MFM-NHPP algorithm in Figure~\ref{fig:randindx}.
It can be readily seen that the Rand index rapidly converges within $1000$ MCMC iterations with reasonable variations, indicating rapid mixing and convergence of the chain. We also notice that in the case where $K = 3$, the rand index can converge to around 0.73 while it can only converge to around 0.66 in the case where $K = 6$, which is consistent with the observations in Figure \ref{fig:cluster_hist}, where the estimation on the number of clusters is more accurate in the case where $K = 3$.  

\subsection{Estimation Performance}\label{sec:est}
We now evaluate the performance MFM-NHPP in terms of estimating the number of clusters, accuracy of clustering (rand index) as well as the estimation of intensity values for each cluster. The two scenarios discussed in previous section are explored. In both scenarios, 100 independent datasets are generated using the steps outlined at the beginning of the section. For each independent dataset, the MFM-NHPP algorithm is run for 5000 MCMC iterations leaving out a burn-in of 2000.   We report the proportion of times the true $K$ is recovered among the $100$ replicates as well as the average rand index estimation among the $100$ replicates. 
\begin{figure}[ht]
\centering
	\includegraphics[width=2 in]{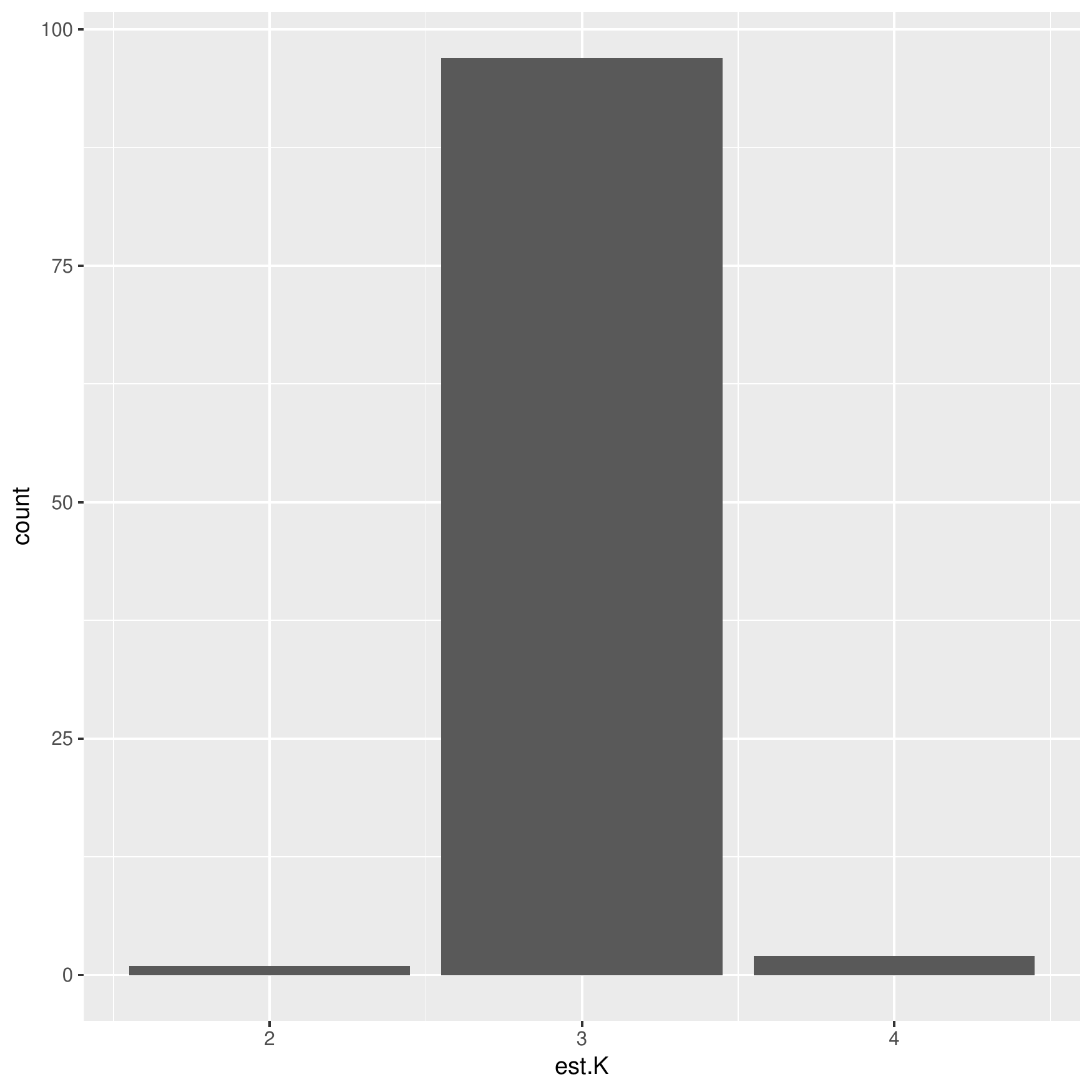}
	\includegraphics[width=2 in]{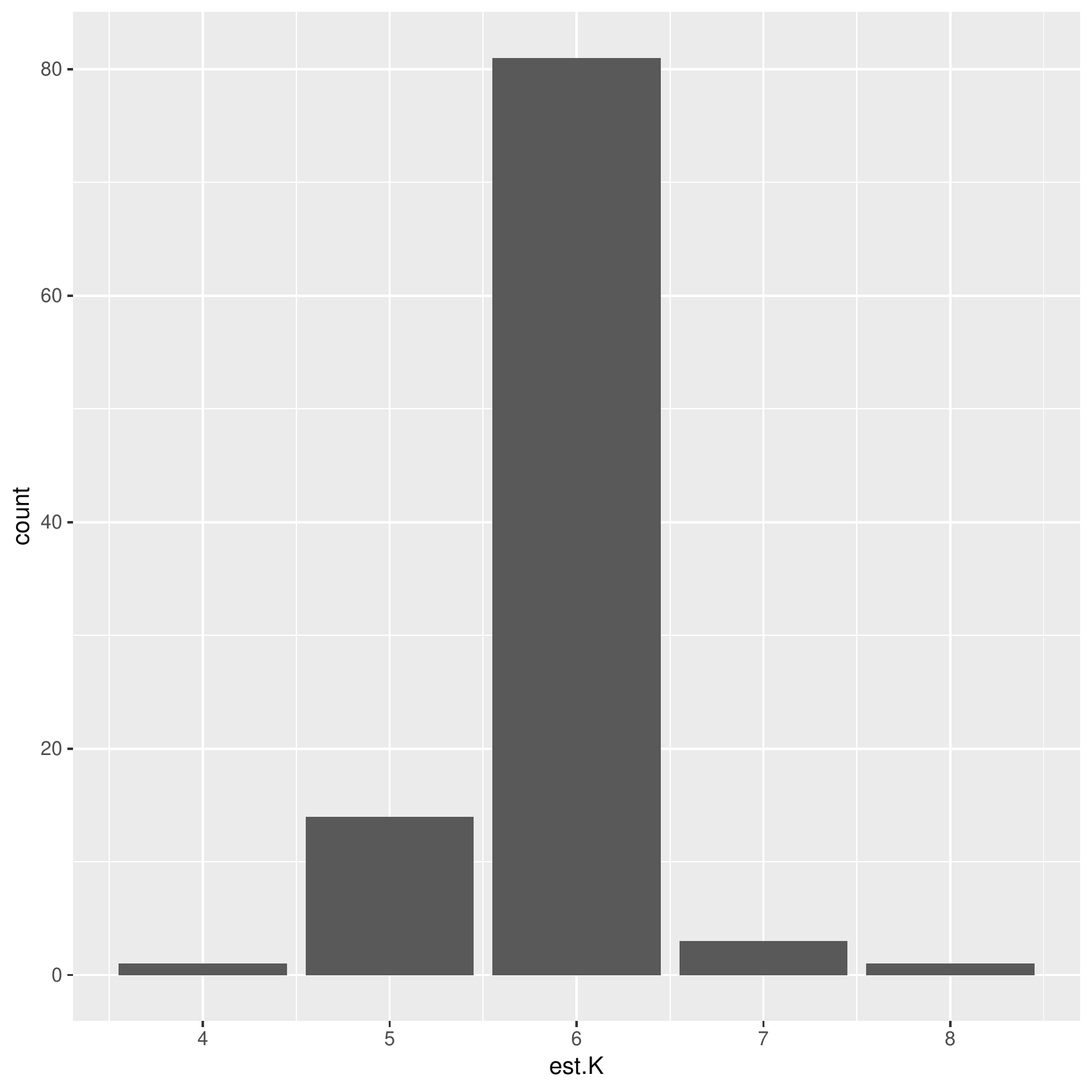}
	\caption{Histograms of estimated number of clusters across 100 replicates for scenario 1 (left) and scenario 2 (right)}
	\label{fig:cluster_hist}
\end{figure}

The summaries of estimating the number of clusters from the 100 replicates are provided in Figure \ref{fig:cluster_hist}. It can be seen from the left panel that when $K=3$, our method can recover the true number of clusters in over $90\%$ of the replicates. From the right panel, we can also see that with large number of clusters like $K=6$, the proposed method can also recover the true number of clusters in over $80\%$ of the replicates.  

\begin{table}
\caption{Summaries using the Dahl's method}\label{tab:1}
\centering
\fbox{%
\begin{tabular}{c|ccccc}
	$K$ & {\bf True } & {\bf Bias} &  {\bf SD} & {\bf MSE} & {\bf RI}\\ \hline
{$K=3$} & 0.2 & 0.0114 & 0.0407 & 0.0018 & {0.730}\\
& 10 & 0.7578 & 0.9460 & 1.4602 &\\
& 20 & -1.2908 & 1.0683 & 2.7960 &\\ \hline
{$K=6$} & 0.2 & 0.0620 & 0.0506 & 0.0064 & {0.662} \\ 
& 5 & -0.1296 & 0.3764 & 0.1571 &\\  
& 20 & 1.0372 & 2.5809 & 7.6703 &\\ 
& 40 & -2.7355 & 2.6416 &  14.3911 &\\   
& 80 & -2.4298 & 8.6088  & 79.2740 &\\
& 200 & -32.2507 & 20.5086  & 1456.5080 &\\ 
\end{tabular}}
\end{table}

\begin{table}
\caption{Summaries using the Posterior Mean}\label{tab:2}
\centering
 \fbox{%
\begin{tabular}{c|cccc}
\hline
$K$ & {\bf True intensity} & {\bf Bias} &  {\bf SD} & {\bf MSE}\\ \hline
{$K=3$} & 0.2 & 0.0174 & 0.0277 & 0.0011 \\
& 10 & 1.2567 & 0.6760 & 2.0316 \\
& 20 & -1.8428 & 0.8402 & 4.0948 \\ \hline
{$K=6$} & 0.2 & 0.0986 & 0.0330 & 0.0108 \\ 
& 5 & -0.2462 & 0.3398 & 0.1749 \\  
& 20 & 1.0081 & 2.4109 & 6.7704 \\ 
& 40 & -2.7069 & 2.2358 & 12.2762 \\   
& 80 & -2.3562 & 8.1082 & 70.6377 \\
& 200 & -33.6250 & 19.8135 & 1519.2920 \\ 
\hline
\end{tabular}}
\end{table}

The accuracy of clustering (rand index) as well as the estimation of intensity values for each cluster using Dahl's method are reported in Table \ref{tab:1}. From those results, it can be seen that intensity values for each cluster are recovered very well in the case when $K = 3$. When $K = 6$, the estimations on intensity values are accurate in most of the clusters except for the cluster with true intensity value of 200. The average random index for both scenarios is around 0.7. And from Table \ref{tab:2}, we see that the estimations of intensity values from two summary methods are consistent in general.

\subsection{Comparison to Competitors}\label{sec:com}

\begin{figure}[tbp]
\centering
	\includegraphics[width=2 in]{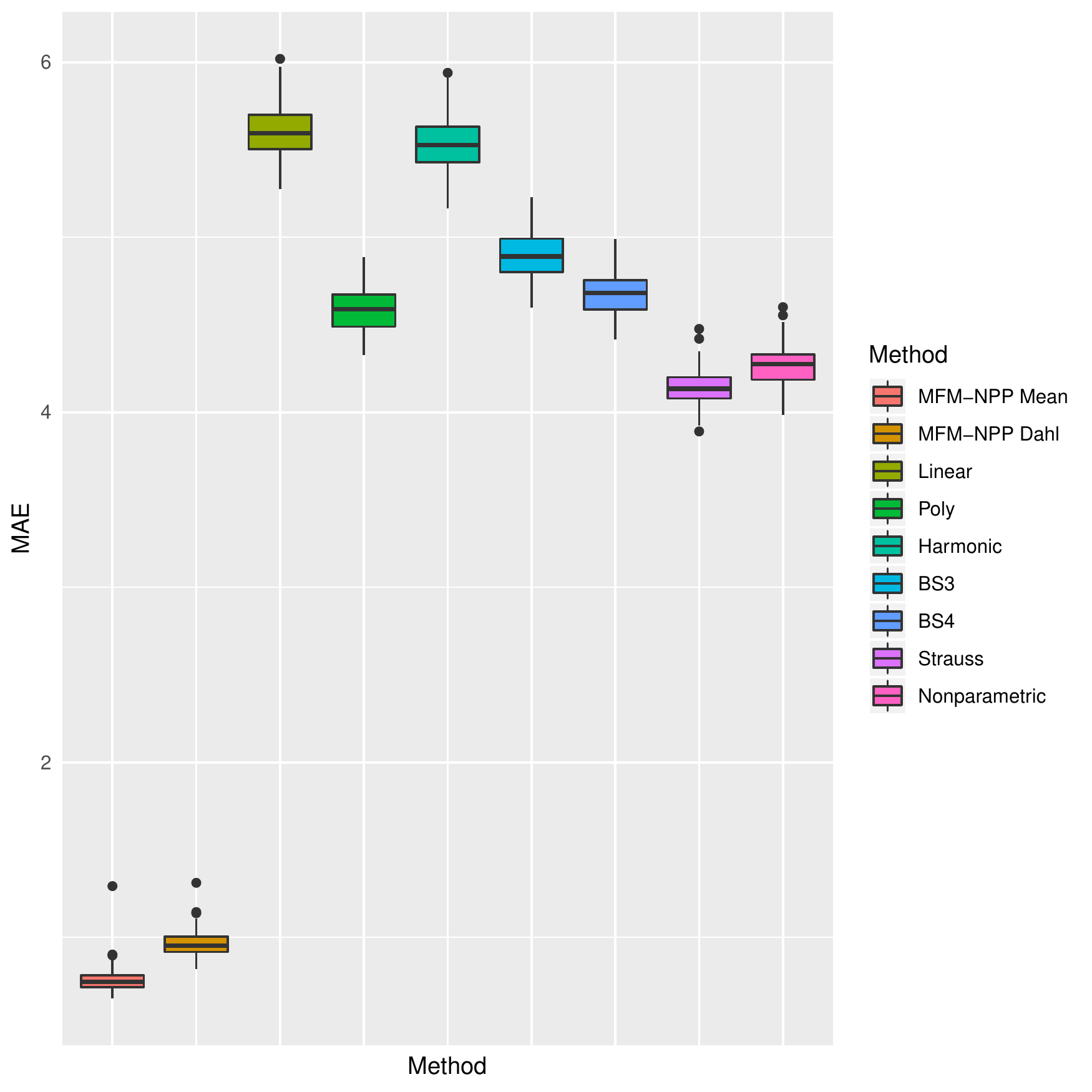}
	\includegraphics[width=2 in]{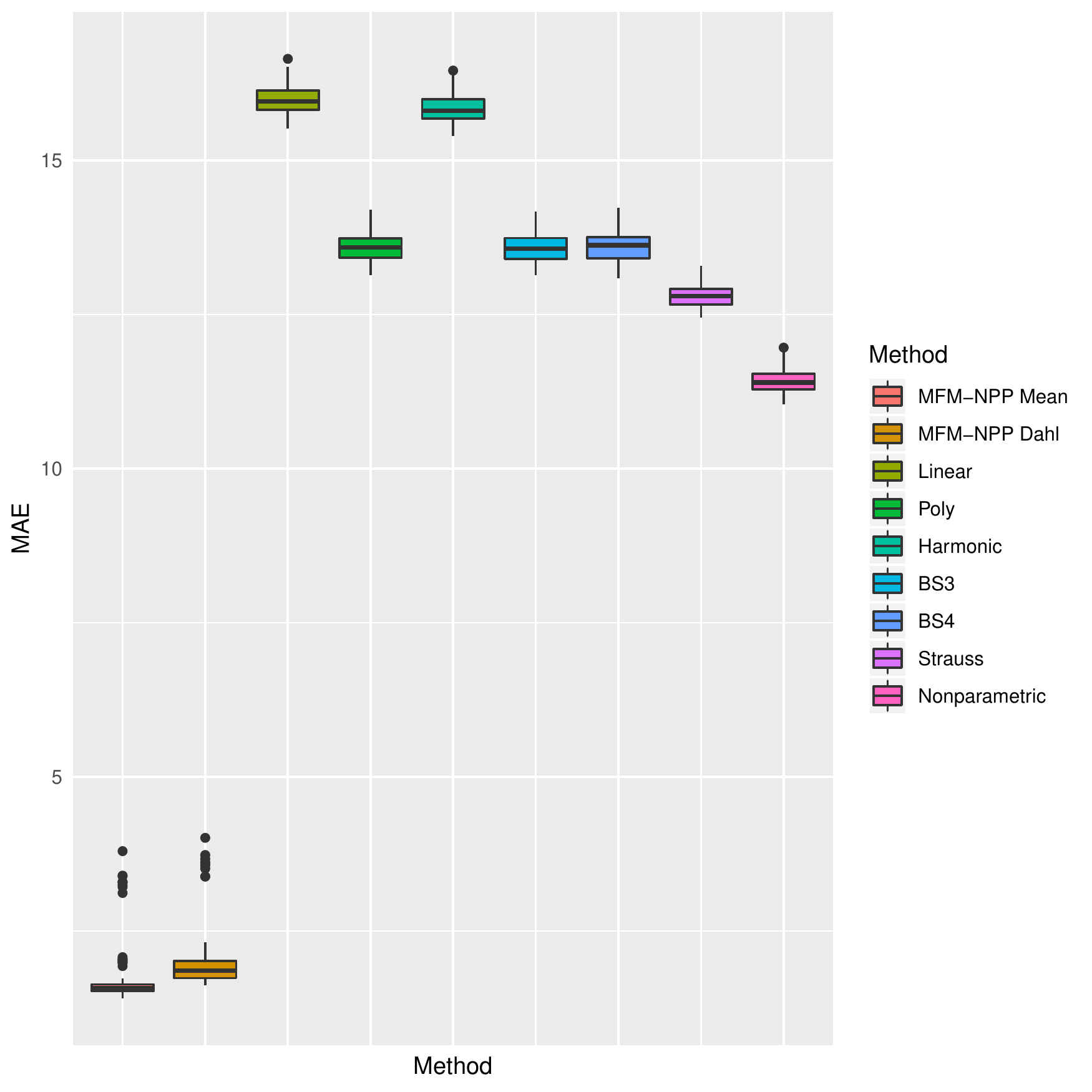}
	\caption{MAE Comparison for scenario 1 (left) and scenario 2 (right)}
	\label{figure_mae}
\end{figure}

In order to compare our methods with other methods, we use the MAE to measure the performance of different methods.  Our method is compared with 7 benchmark methods (Poisson Process with linear tread of the coordinates $x$ and $y$;  with Polynomial of order 3 in the coordinates $x$ and $y$; Poisson Process with Harmonic polynomial of order 2 in the coordinates $x$ and $y$; Poisson Process with 3 degree of freedom and 4 degree of freedom B-splines in the coordinates $x$ and $y$; Strauss process with linear tread of the coordinates $x$ and $y$; nonparametric kernel estimation of  Poisson Process) in \textbf{spatstat} \citep{baddeley2005spatstat}. The boxplot of the MAE of our method and the 7 competitors in two simulation scenarios are shown in Figure~\ref{figure_mae}. From those plots, we can see that our methods clearly outperform the competitors in the MAE comparison in both scenarios. The results in Figure \ref{figure_mae} indicate that our proposed methods (summarized by Dahl's method \ref{sec:sum_mcmc} and posterior mean) have better overall intensity estimation than other seven methods in both scenarios.

\section{Analysis of USGS Data}\label{sec:real_data}
\begin{figure}[h]
\centering
 \includegraphics[width= 2in]{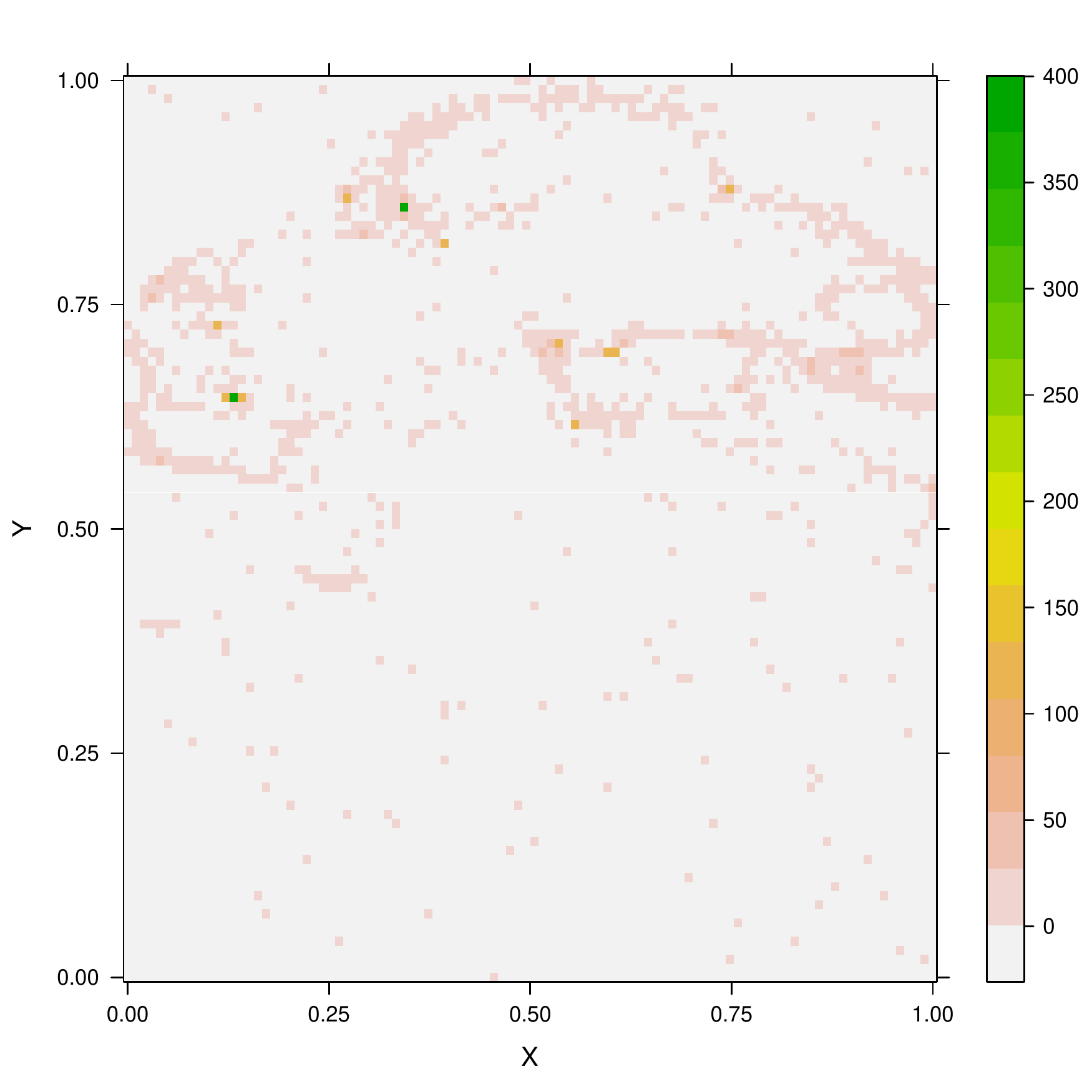}
 \includegraphics[width= 2in]{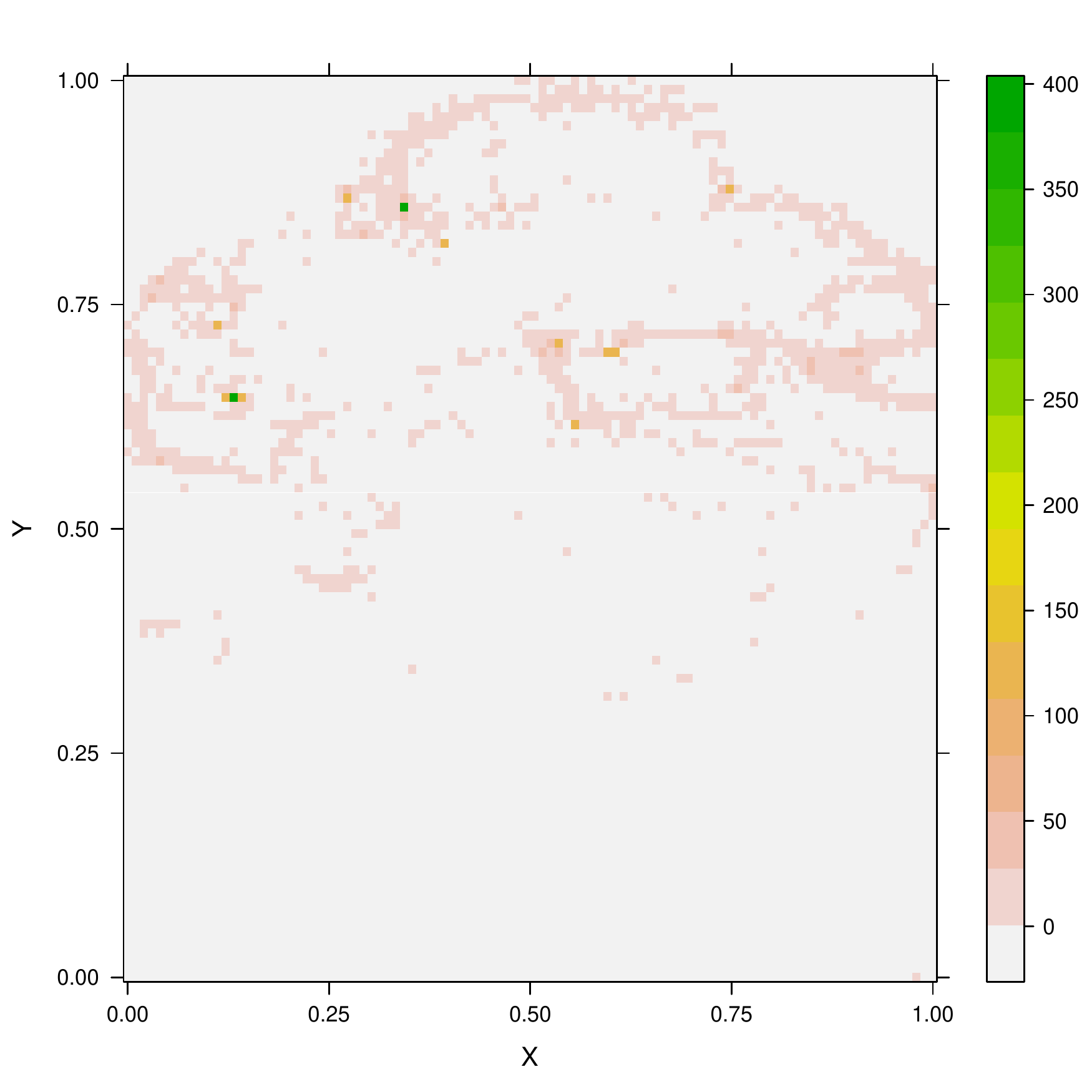}
	\caption{Intensity Estimation for $100\times 100$ resolution given by 
	the Dahl's Method (left) and the posterior mean (right)}
	\label{fig:real_data_intensity}
\end{figure}

In this section, we present a detailed analysis of USGS data based on our proposed method and try to explore the heterogeneity of the earthquake activities over the space. Based on the model in \eqref{eq:MFMNHPP}, we divide the spatial domain in three different pixel resolutions ($20\times20$, $50\times50$, and $100\times100$). We employed Algorithm \ref{algorithm} with $n = 400, 2500, 10000$, $\gamma =1$ and $a=b=1$ to fit the MFM-NHPP model; and we refer to this as the MFM-NHPP algorithm.  A truncated Poisson prior with mean $1$ is assumed on $k$. A total of 15,000 MCMC samples are saved after 5,000 burn-in. The MAE values of our methods and 7 benchmark methods as mentioned in Section \ref{sec:com} are report in Table \ref{table:real_data_comparision}.
\begin{table}
	\caption{\label{table:real_data_comparision}MAE Comparison for Real Data ((1): MFM-NHPP Posterior Mean; 
	(2): MFM-NHPP Dahl's Method; 
	(3)-(9): benchmark competitors)}
\centering
\fbox{
	\begin{tabular}{cccccccccc}

	&(1)&(2)&(3)&(4)&(5)&(6)&(7)&(8)&(9)\\
	\hline
	$20\times 20$&3.86 &5.32&   27.9&29.7&     28.5& 29.7&29.8&    22.1&          29.3\\
	$50\times 50$&0.593&0.832&  4.85&  5.05&  4.92& 5.05& 5.06&   4.10&5.04\\
	$100\times 100$ &0.182& 0.265&    1.28& 1.30&      1.29& 1.31& 1.31&   1.14&  1.31\\
	\hline
	\end{tabular}}
\end{table}
From the results in Table \ref{table:real_data_comparision}, we see that our proposed methods have consistently better intensity estimation than other 7 methods. Furthermore, we compared the LPML values based on \eqref{eq:lpml_estimate}. The LPML of three different resolutions ($20\times 20$, $50\times 50$, and $100\times 100$) are 68970, 75116, and 79148, respectively. We see the resolution with $100\times 100$ has the best estimation performance based on LPML. The estimated number of the clusters based on Dahl's method is 8. The estimated intensities of each cluster are 0.015, 1.590, 1.614, 6.560, 16.677, 38.574, 125.863, and 374.011. The numbers of the area in each cluster are 9083, 96, 524, 196, 66, 23, 10, and 2, respectively.  The estimated intensity plots based on Dahl's method and posterior mean are shown in Figure \ref{fig:real_data_intensity}.

\begin{figure}
\center
	\includegraphics[width=3 in]{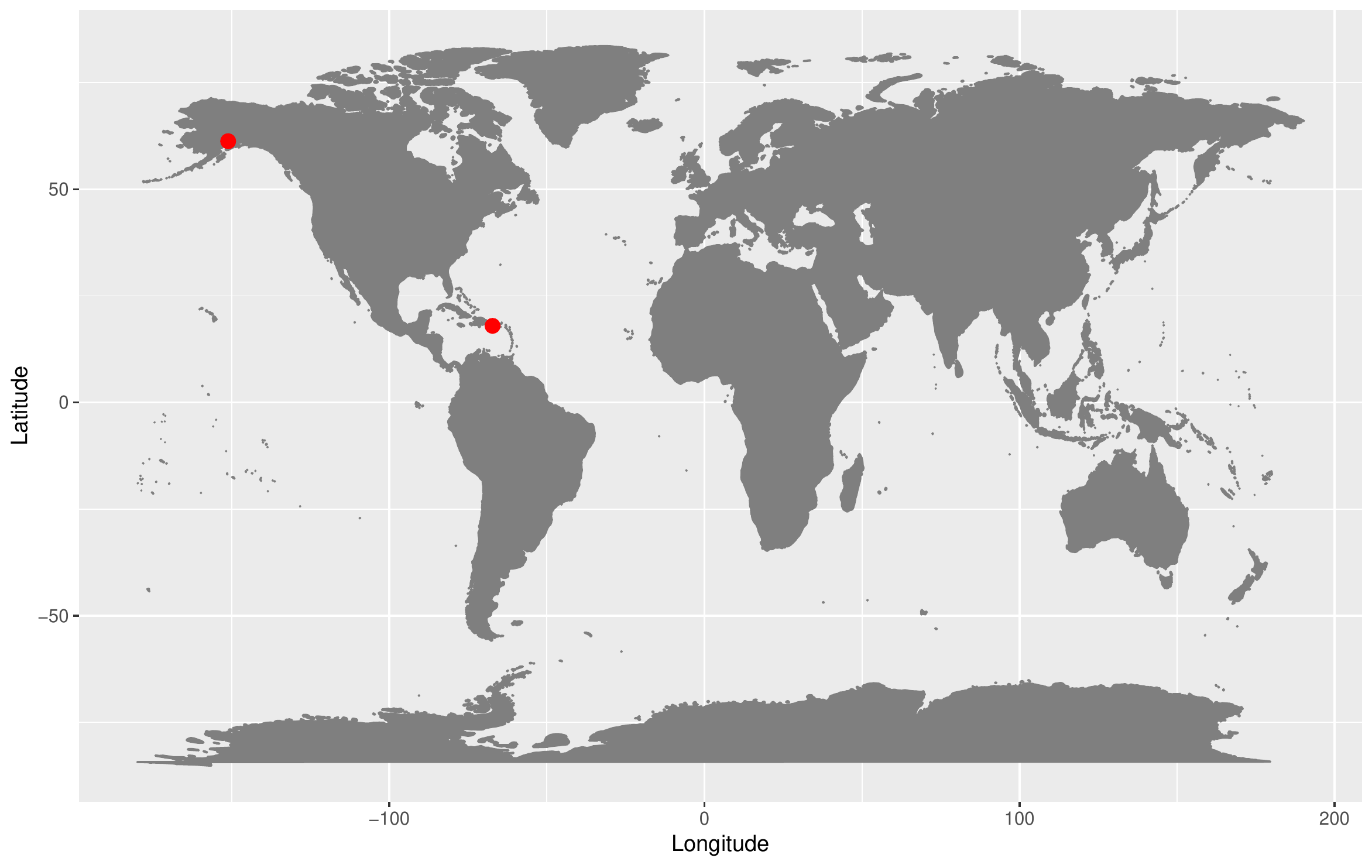}
	\caption{Locations with the highest intensities.}
	\label{fig:hot_points}
\end{figure}

From the intensity plots and clustering results, we see that there are two areas with very high intensity for earthquakes. Nearly 90\% earthquakes will occur in just 5\% areas around world. In 90\% region around the world, there is almost no earthquake occurrence. The approximated locations of two region belongs to highest intensity cluster is $(-151.15,61.28)$ and $(-67.23,17.96)$ which are near to Alaska's central coast and Puerto Rico trench, respectively. These results are consistent with seismic zone analysis \citep{mccann1985earthquake,kelleher1970space}.  The two locations are shown in Figure \ref{fig:hot_points}. 
\section{Discussion}\label{sec:discussion}
In this paper, we develop a nonparametric Bayesian intensity estimation for nonhomogenous Poisson process based on Mixture of Finite Mixtures model. This statistical framework was motivated by the heterogeneity of earthquake activities over the space.  

Our simulation results indicate that the proposed method can recover the heterogeneity pattern on intensity over space, and obtain better intensity estimations than other intensity estimation methods of Poisson process. Illustrated by the analysis of USGS data, our nonparametric intensity estimation has better performance than other models under Poisson assumptions by revealing the heterogeneity pattern of the earthquakes' occurrences. 

In addition, three topics beyond the scope of this paper are worth further investigation. First, we need to add spatially dependent structure on the intensity of each small areas. Second, spatial covariates are not taken into consideration. In the future, adding spatially dependent covariates is desirable for the potential improvement in intensity estimation. Building a nonparametric Bayesian model beyond the Poisson assumption is also devoted to future research.

\section*{Acknowledgement}
Dr. Hu's research was supported by Dean's office of College of Liberal Arts and Sciences at University of Connecticut.

\appendix

\section{Full Conditional Distributions} \label{FCD}

The full conditional distributions in Markov chain Monte Carlo (MCMC) sampling algorithm \ref{algorithm} in Section \ref{sec:bayes_comp} are given as follows. 

For each term $\lambda_r$ in $\lambda = (\lambda_1, \ldots, \lambda_k)$, the full conditional distribution is:

\begin{equation}
\label{eq:lambda}
\begin{aligned}
    p(\lambda_{r} \mid N,z)  &\propto  \mbox{Gamma}(\lambda_{r}) \prod_{z_i = r} \mbox{Poisson}(N(A_i);\lambda_{r})\\
    &\propto \lambda_{r} ^{a-1} e^{-b\lambda_{r}} \prod_{z_i = r} \lambda_{r} ^{N(A_i)} e^{-\lambda_{r}}\\
    &\propto \lambda_{r} ^{\sum_{z_i=r} N(A_i) + a -1} e^{-\lambda_{r}(\sum_{i=1}^{n}I(z_i=r)+b)}\\
    &\propto \lambda_{r} ^{\bar{N}_{r} + a -1} e^{-\lambda_{r}(n_{r}+b)}\\
\end{aligned}
\end{equation}

This implies that $p(\lambda_{r} \mid N,z) \sim \mbox{Gamma}(\bar{N}_{r}+a,n_{r}+b)$. For each term $z_i$ in $z = (z_1, \ldots, z_n)$, the full conditional distribution is:
\begin{equation}
\label{eq:z}
\begin{aligned}
   \text{If } c = c_j \text{ for some } j \neq i, p(z_{i} = c \mid z_{-i}, N, \lambda)  &\propto  p(z_{i} = c \mid z_{-i}) \mbox{dPoisson}(N(A_i);\lambda_c) \\
   \text{If } c \neq c_j \text{ for all } j \neq i, p(z_{i} = c \mid z_{-i}, N, \lambda)  &\propto  \frac{V_n(\abs{\mathcal{C}_{-i}}  +1)}{V_n(\abs{\mathcal{C}_{-i}}} \gamma  m(N(A_i))\\
\end{aligned}
\end{equation}
where
\begin{equation}
\label{eq:mni}
\begin{aligned}
    m(N(A_i))  &=  \int \mbox{Gamma}(\lambda) \mbox{Poisson}(N(A_i);\lambda) \mbox{d} \lambda\\
    &= \int \dfrac{b^a}{\Gamma(a)} \lambda ^{a-1} e^{-b\lambda}  \dfrac{\lambda ^{N(A_i)} e^{-\lambda}}{N(A_i)!}\mbox{d} \lambda\\
    &= \dfrac{b^a \Gamma(N(A_i)+a)}{\Gamma(a)(b+1)^{N(A_i)+a}N(A_i)!} \int \dfrac{(b+1)^{N(A_i)+a}}{\Gamma(N(A_i)+a)}\lambda ^{N(A_i) + a-1} e^{-(b+1)\lambda} \mbox{d} \lambda\\
    &= \dfrac{b^a \Gamma(N(A_i)+a)}{\Gamma(a)(b+1)^{N(A_i)+a}N(A_i)!}
\end{aligned}
\end{equation}
\bibliographystyle{chicago}

\end{document}